# CALCITE, VATERITE AND ARAGONITE FORMING ON CEMENT HYDRATION FROM LIQUID AND GASEOUS PHASE

*Ewa T. Stepkowska* [1], *J. L. Pérez-Rodríguez* [2*], *M. J. Sayagués* [2] and *J. M. Martínez-Blanes* [2]

[1]Institute of Hydroengineering PAS, ul. Koscierska 7, 80 953 Gdañsk, Poland
[2]Instituto de Ciencia de Materiales de Sevilla, CSIC, c/Americo Vespucio, s/n, 41 092 Isla de la Cartuja, Sevilla, Spain

## Abstract

Cement hydration products were studied as influenced by the hydration conditions (hydration time in liquid phase; relative humidity, RH, in gaseous phase). The formation of calcium hydroxide (portlandite, P) and its transformation to calcium carbonates is mainly discussed here.

More hydration products, including P, were formed in liquid phase (paste) than in water vapor (powder), due to the higher availability of water molecules. Full hydration was observed only in the paste hydrated for 6 month, otherwise the P content, estimated from its water escape, $\Delta M(400–800°C)$, increased after storage in water vapor of the prehydrated paste.

All the three polymorphs of $CaCO_3$ (calcite, vaterite and aragonite) were found on prolonged contact with air of the hydrated powder (XRD, HRTEM). Their content was dependent on sequence of RH conditions on hydration: higher after water retention, WR, on lowering RH=1.0→0.95→0.5, than after water sorption, WS, on increasing RH in the inverse order. It increased also on wetting and drying, both of hydrated powder and paste. Ca was found to accumulate on the micro-surfaces of WR samples (SEM, TEM), whereas more Al was observed on WS samples and the crystallinity of hydration products was here higher (ED). Dissolution-diffusion-recrystallization was possible: small Al-ions concentrated at one end and the bigger Ca ions - at the other end of some needles (TEM).

At 400–500°C the P in cement transforms in air into $CaCO_3$, which decomposes at 600–700°C. Thus the sensitivity to carbonation was estimated from $\Delta M(600–800°C)$. This value was similar in pastes hydrated for 1 month and in powder (WR). It was lower in powder WS and much lower in the paste (6 months). It increased pronouncedly when the prehydrated paste was stored in water vapor in WS.

The nanocrystals of portlandite, vaterite and aragonite, embedded in the amorphous matrix, were observed by HRTEM in the hydrated powder. They may contribute to the cement strength.

## Introduction

In a previous study it was suggested to evaluate the cement quality by its hydration from the gaseous phase [1–2]. Here the hydration products from liquid and gaseous phase are compared, finding both qualitative and quantitative differences and getting some information on the hydration process. Most attention is paid to the variability in portlandite (P) content, in the sensitivity to carbonate formation (carbonation) and in the relative content of calcite, vaterite (V), and aragonite (A), which developed on storage in air of the hydrated cement and were identified by XRD. Calcite is usually present in the superficial layer of the hydrated cement (mortar), V and A are not frequent, but their presence was also reported (see below). Sensitivity to carbonation was estimated from the mass change on heating to elevated temperatures. The microstructure was studied by SEM, TEM and HRTEM and the influence of some microstructural parameters is discussed.

---

\*     Author for correspondence: E-mail: jlperez@cica.es

Not only the availability of water molecules from liquid or gaseous phase is different, but also the microstructure of the hydrating cement, i.e. powder in water vapor and paste in liquid water. The microstructural features of the cement-water interaction products evoke a high interest [3–5]. For example, the Mg:Ca ratio and the salinity influences the form of calcium carbonate minerals (spherulite, radial fibrous and concentric oolites, prismatic and granular crystals), their crystallinity decreasing with the increase in this ratio [6–7]. Hydration products are dependent also on the age of cement powder and on hydration time, which is discussed below.

*V and A in the hydrated cement and in its components*

Formation of V and A is triggered by as yet unknown mechanism [8–9]. They were found by Cole and Kroone [10–11] in mortars and in calcium silicate hydrate on their carbonation (reaction with carbon dioxide), formed in the following hydration–carbonation steps: anhydrous cement minerals→siliceous residue+calcium hydroxide→poorly crystallized (vaterite+aragonite+calcite)→well crystallized calcite and some portlandite crystallites.

V occurs either as hexagonal plates or as 'spherulitic formations' of a lower crystallinity [12–13], whereas A is orthorhombic.

V may develop with time in a hot climate enhanced by the presence of combustion gases [14]: it was observed in the bonding mortars of marble inlays of Florence Cathedral [15], in San Pedro sculpture at 'Puerta del Perdon' of Sevilla Cathedral [16] and it formed from larnite ($Ca_2SiO_4$) in the calcium silicate hydrogel crystals in presence of $CO_2$ [17]. The development of V is due to the carbonation of the high lime hydrogel. The presence of the imperfectly crystalline $Ca(OH)_2$, the similarity of the crystal system (hexagonal) and the nearly double hexagonal parameter *a* of the crystal lattice of V as compared to portlandite (7.1473 and 3.5899 A, respectively) favor the layer type structure of vaterite [17].

*The influence of temperature on formation of calcium carbonate polymorphs*

The precipitate of calcium carbonate from a supersaturated solution is initially predominantly amorphous transforming to the metastable polymorphs. These change finally to stable calcite, which is forming preferably at a low temperature and is decreasing with the temperature increase. The vaterite and calcite content is equal at

30°C and V predominates at 40°C, when also A starts to form and it is the major constituent at 60–80°C. At any temperature all the crystals change finally to calcite due to dissolution of the metastable form. The growth rate of calcite crystals is proportional to the number of growth sites, i.e. to the surface area of calcite, which is very rough and active if created by crushing. The surface of calcite crystallites is negatively charged, whereas that of V is positive, both attracting counterions [18].

Dry heating for a few minutes at 400°C causes a transformation of metastable polymorphs into calcite as the stable one [17].

*Influence of solubility on transformation of carbonate polymorphs*

The higher solubility of the meta-stable polymorphs, increasing in the sequence: calcite<A<V, induces their change to calcite at any temperature in solution.

The enthalpy of dissolution in $CO_2$–$H_2O$ system at 298 K is 10.83 and 15.79 kJ mol$^{-1}$ for calcite and V, respectively [19]. The solubillity product at 25°C is: $\log K_{sp}$= –8.48; –8.34 and –7.91 for calcite, A and V, respectively [18].

*Transformation of A to calcite*

The transformation of A to calcite is slightly endothermic (about 85 cal mol$^{-1}$) and around the transformation temperature, $T_{tr}$ a decrease in crystallite size and in unit cell volume occurs, which indicates a disordering of the crystal lattice. Wolf and Guenther [20] indicate $T_{tr}$=455±10°C at the transition enthalpy 403±8 J mol$^{-1}$ and activation energy 370±10 kJ mol$^{-1}$ (452±19 kJ mol$^{-1}$, according to Davies *et al*. [21] and 93–95 kcal mol$^{-1}$ according to Subba Rao [22]). The specific density is 2.93 and 2.71 g cm$^{-3}$, respectively. Nucleation (not diffusion) is the rate determining step. The transformation may occur as follows [22]:

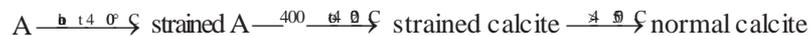

Some data indicate that Mg-ions cause a reduction in activation energy and an increase in the rate of transformation. A higher Na-content lowers slightly the temperature and activation energy of transformation, but its energy is higher [22]. A is the high pressure form of calcium carbonate [23].

The well crystallized calcite decomposes at 850°C (nominally at 898.6°C) indicating a strong endothermic effect in DTA. Only a slight effect at 700°C is observed on decomposition of the poorly crystallized species [10–11]. In cement this reaction occurs at 600–700°C [24].

*Transformation of V to calcite*

Hexagonal V transforms into calcite on heating to temperatures higher than 400°C [25]: by 25% after 4h at 463°C and by 91% at 485°C [22]. The reaction is slightly exothermic (about 145 cal mol$^{-1}$), the activation energy being 92–102 kcal mol$^{-1}$ and the nucleation process creating the rate determining step. A large decrease in the crystallite size occurs around the transformation temperature but the change in unit cell volume is negligi-ble [22]. This hexagonal V was synthesized by addition of potassium carbonate to the supersaturated calcium chloride solution in presence of sodium hexametaphosphate.

Wolf and Guenther [20] indicate the thermodynamic parameters of this reaction as: $T_{tr}$=320–460°C, depending on the way of preparation, at transition enthalpy –3.2±0.1 kJ mol$^{-1}$ and activation energy 250±10 kJ mol$^{-1}$.

*Examples of occurrence in nature*

Aragonite (named after Aragon, Spain) is very common in biological and rare in non-biological material [20]. It was found in the near surface deposits formed at a low temperature. It occurs in the marine environment (possibly due to the abundance of Mg ions in solution), e.g. it was found in the sea bed (18 to 100 m) off the edge of the N-West shelf of the Western Australia [26], in the bottom sediments of the Adriatic Sea, along the Pacific coast of North America. It crystallized from sea water in the floor of Atlantic and Indian Oceans. Calcium carbonate in pearls occurs usually in form of aragonite [23].

Vaterite occurrence in nature is rare in biological and very rare in non-biological material [17, 20, 23]. Precipitation of V was observed in drilling fluids in wells of USA and New Zealand. It was found in Ordovician shales in the Appalachians of Quebec [14, 27]. It was identified in Holkham Lake in North Norfolk [28] but it transformed into calcite indicated by re-examination of this water [8].

Vaterite was found as a major constituent in a carbonated calcium silicate hydrogel complex formed of larnite (belite) at Ballycraigy, Larne, Northern Ire-　land [17]. Larnite is one of the main components of cement.

Vaterite was formed from carbonate rocks, dissolved probably by the natural gas [14]. This may also occur in stone monuments in places of high traffic.

It was identified in hard tissues of some organisms, in molusc shells, in human gallstones, in otoliths of some fish, in zones of thermal methamorphism [29].

## Materials

Ordinary portland cements (OPC) of grade 43 [Indian Standards IS:8112-1989] was supplied by National Council for Cement and Building Materials (NCB, New Delhi, India), of the chemical composition: CaO=61.0%, SiO$_2$=20.9%, Al$_2$O$_3$=5.3%, Fe$_2$O$_3$=3.1%, MgO=3.6%, K$_2$O=0.89%, SO$_3$=1.5%, Na$_2$O=0.45%, LOI=2.7%.

These values as measured by EDX at magnification ×500 on the mortar hydrated for 1 month and acetone treated were: CaO=59.6–53.4$^*$%, SiO$_2$=28.5–32.5$^*$%, Al$_2$O$_3$=4.7–5.9$^*$%, Fe$_2$O$_3$=3.9–4.0$^*$%, MgO=2.6–3.6$^*$%, K$_2$O=0.68–0.71$^*$% ($^*$selected surface not containing Portlandite crystals).

The parent cement powder indicates the standard composition of Ordinary Portland Cement OPC, Taylor, [30]: XRD peaks of alite (C$_3$S=Ca$_3$SiO$_5$), belite (larnite, C$_2$S=Ca$_2$SiO$_4$) and ferrite (brownmillerite, Ca$_2$(Al, Fe)$_2$O$_5$) were found. Alite and belite grains were observed by SEM (Fig. 1a–b). Some gypsum XRD peaks occurred in selected samples. No portlandite, P, vaterite, V, or aragonite, A, peaks were found.

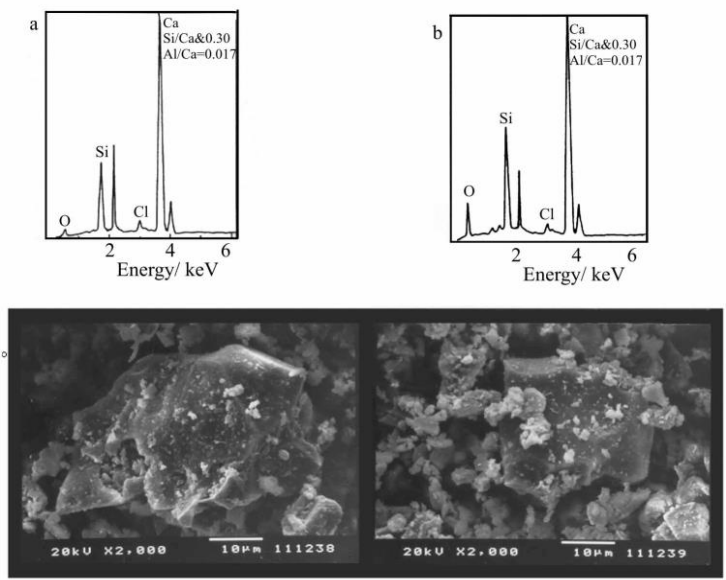

**Fig. 1** Alite (a, 238) and belite (b, 239) in cement powder (C-43) at magnification ×2000. The EDX spectrum shows the ratio Ca:Si=3.33≅3 (a, 238) at a high magnification and 2.17≅2 (b, 239) at a point acquisition

After wetting and drying a slight increase in the relative intensity occurred, both of the joint alite+calcite peak at 3.034 Å and of the joint alite+belite peaks at 2.74 and 2.61 Å. Thus a slight hydration of alite to belite and some carbonation of calcium hydroxide occurred.

## Methods

### Hydration in water vapor

Samples of cement powder and of hydrated pastes were stored at the given relative humidity, RH, at 30°C=const. and at atmospheric pressure, either increasing successively, defined as WS (RH=0.5→0.95→1.0, or WSo if predried at 110°C) or decreasing successively in WR (RH=1.0→0.95→0.5, 2 weeks in each case [31]).

### Hydration in liquid water

Series I. Paste samples to compare with powder samples were prepared by mixing the cement powder with distilled water, at the water/cement ratio $w/c$=0.4. Two cubes 3×3×3 cm (PI and PII) were formed by hand, stored for 28 days in water submerged state at room temperature and air dried.

Series II. Standard paste (mortar) was prepared at tap water/cement ratio, $w/c$=0.4, formed into standard cube (at NCB) and stored at room temperature in water submerged state for 1 month and for 6 months. Sample fragments were (i) air dried, a.d. (ii) acetone treated and air dried, ac. Additionally they were wetted and dried at 40°C for XRD; multiple similar processes occur on weathering.

*Static heating*

The hydrated cement powder and paste samples were heated statically both before (defined as SH) and after the sorption test (defined as WS and WR). The heating temperatures were: 110°C overnight, 220°C for 8 h, 400°C for 4 h, 600°C for 2 h and 800°C for 1 h. After heating at 220°C samples were transferred from aluminium to ceramic containers.

Both the water content (W, escaping at 110°C as EV) and the mass change on heating at the given temperature $T$ [in °C], i.e. $\Delta M(T)$ were calculated in mass% in relation to the sample mass at 800°C, i.e. $\Delta M(800°C)$=0 and $\Delta M(110°C)$=non-EV.

The standard deviation in case of pastes hydrated for 1 month was for W:±0.02 to 0.3% and for cement powder, hydrated in water vapor it was ±0.06 to 0.4%. The remaining values of standard deviation are indicated in Table 1b.

The static heating was done directly after the cement hydration, the XRD, SEM and TEM study was done after about 1 year.

*Components of the cement hydration products*

Water evaporable at 110°C (EV) is proportional to the specific surface [31].

The non-EV [$\Delta M(110$–$800°C)$ with $\Delta M(800°C)$=def. 0], depends on the content of the hydration products [31], i.e.

• $\Delta M(110$–$220°C)$ is water bound with low energy, e.g. water in hydrates or as dipolar molecules.
• $\Delta M(220$–$400°C)$ is the 'zeolitic' water, bound in the poorly crystalline CSH-gel,
• $\Delta M(400$–$800°C)$ represents approximately water released from portlandite, on its dehydroxylation, P($H_2O$), if no carbonates are present below 400°C. The nominal temperature of P decomposition is 580°C, but in cement it occurs either at 400°C, [1], [32, 33], or at 400–500°C [34], depending on cement quality, on its 'age' and on heating time. If $\Delta M(400°C)>\Delta M(220°C)$, the carbonation proceeded at or below this temperature [1].
• $\Delta M(600$–$800°C)$ represents the sensitivity to carbonation, i.e. the content of $CaCO_3$, which formed of P either on prolonged contact with air, or after heating at 220°C and contact with air [1], or at or above 400°C, when also all the metastable polymorphs transform into calcite [17]. This transformation and its temperature depends on the crystallinity of Ca(OH)$_2$, on crystallite size, on heating time and on density of the matrix. The $CaCO_3$ decomposes nominally at 898.6°C, but in cement generally at 600–700°C [24].

**Table 1a** Results of static heating of cement powder and cement paste (series I) in mass% *vs.* the mass at 800°C

| Sample | Test | EV (110°C) | non-EV | ΔM (220°C) | Rehydration | ΔM (400°C) | ΔM (600°C) | Note |
|---|---|---|---|---|---|---|---|---|
| a | b | c[*] | d | e | f[**] | g | h | i |
| Powder | WS | 10.5 | 9.1 | 6.8 | 7.0 | 4.8 | 1.9 | water vapor |
| C-43 | WR | 15.5 | 10.4 | 7.6 | 7.7 | 4.5 | 3.0 | |
| Paste PI | SH | 17.3 | 13.4 | 11.2 | – | 7.8 | 3.4 | |
| 1 month | WS | 20.8 | 17.7 | 15.6 | 24.9 | 16.4 | 7.4 | liquid water |
| (C-43) | WR | 22.6 | 18.6 | 14.4 | 23.8 | 18.0 | 3.9 | |
| Paste PII | SH | 15.7 | 13.2 | 9.7 | – | 7.2 | 2.7 | |
| 1 month | WS | 18.0 | 17.3 | 13.4 | 22.2 | 14.1 | 6.2 | liquid water |
| (C-43) | WR | 18.9 | 18.3 | 14.2 | 23.0 | 14.2 | 2.9 | |

**Table 1b** Results of static heating of cement paste (series II) in mass% *vs.* the mass at 800°C [Stepkowska *et al.*, 2002]

| Sample | Test | EV (110°C) | non-EV | ΔM (220°C) | Rehydration | ΔM (400°C) | ΔM (600°C) | Note |
|---|---|---|---|---|---|---|---|---|
| a | b | c[*] | d | e | f[**] | g | h | i |
| Paste | SH | 13.2 | 15.6 | 12.82 | – | 8.92 | 3.3 | |
| C-43 | WS | 9.9 | 17.2 | 14.38 | 16.4 | 9.19 | 0.007 | tap water |
| 1 month | WR | 14.9 | 17.1 | 14.01 | 16.3 | 11.40 | 0.225 | |
| Paste | SH | 14.9 | 19.2 | 15.4 | – | 10.73 | 0.78 | |
| C-43 | WS | 12.2 | 18.8 | 13.6 | 23.0 | 10.46 | 6.03 | tap water |
| 6 months | WR | 15.3 | 20.7 | 15.3 | 24.2 | 10.81 | 2.83 | |
| Stand. | 1m | 0.02–0.3 | 0.02–0.1 | 0.03–0.3 | | 0.05–0.3 | 0.003–0.6 | |
| dev. | 6m | 0.02–0.8 | 0.06–0.9 | 0.07–0.8 | | 0.2–1.7 | 0.04–1.8 | |

[*]W(RH=0.5) – ΔM(110°C);
[**]increase in mass after transfer from aluminium to ceramic crucibles, thus on contact with air. The difference between pastes, series I (1m), [PI and PII] and paste series II (1m) consists in (i) 'younger' cement powder by about 1 year, (ii) hand compaction in series I, resulting in lower density as compared with standard compaction, series II, (iii) distilled water, series I, *vs.* tap water, series II. Here SH means static heating of hydrated pastes before their successive storage in water vapor in WS or WR

X-ray diffraction (XRD) was studied by the Siemens Kristalloflex equipped with graphite monochromator, CuK$_\alpha$ and the computer SICOMP PC 16-20 with files of standard minerals JCPDS-ICDD. Paste samples were prepared by crushing not grinding (except one control sample).

Scanning electron microscopy (SEM) was done by the JEOL JSM-5400 Scanning Electron Microscope, equipped with Energy Dispersive X-ray Analysis detector (EDX) on a surface prepared by sample splitting.

High resolution transmission electron microscopy (HRTEM) was done by the Philips CM200 microscope, working at 200 kV, with a side entry goniometer, equipped with an EDX detector. Generally only a very little radiation damage was observed when focusing the beam on the sample. Several small and thin particles of the powder, hydrated in WR or WS were suspended in acetone, dropped on a copper grid, dried, coated with carbon film and selected for the TEM study and further analysis.

## Results and discussion

Results of static heating of cement hydrated (i) in water vapor (powder) and (ii) in liquid water (paste) are compared in Table 1a, series I. Hydration time in both cases was the same, i.e. (i) 4 weeks in water vapor at a high relative humidity, i.e. 2 weeks at RH=0.95 and 2 weeks at RH=1.0, both at 30°C and (ii) 4 weeks in liquid (distilled) water at room temperature; in such a system 0.95<RH<1.

XRD traces of the cement powder hydrated in WS or in WR, are shown in Figs 2 and 3, respectively. Histograms of the relative intensity I of the most important XRD peaks of portlandite (P, Ca(OH)₂) and of calcium carbonates (A, V and calcite) are shown in Figs 4 and 5.

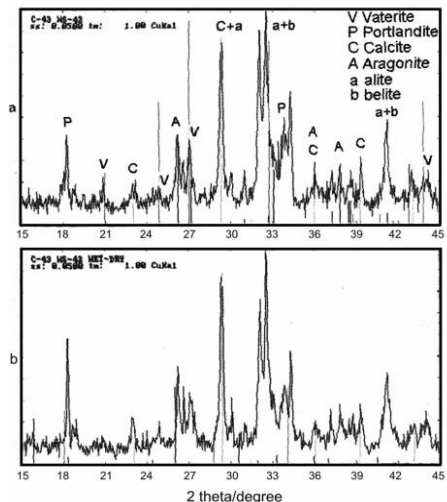

**Fig. 2** XRD patterns of cement powder C-43 hydrated in water vapor at increasing RH (WS) (a), followed by wetting and drying at 40°C (b). Main calcite peak at 3.03 Å coincides with alite (Ca$_3$SiO$_5$) peak at 3.04 Å

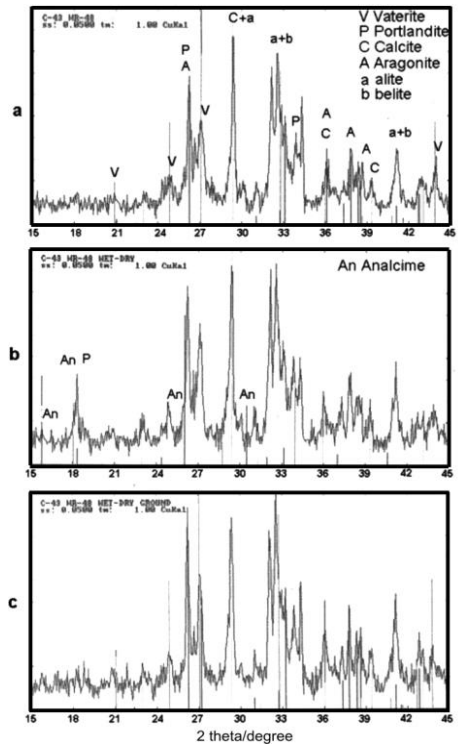

**Fig. 3** XRD patterns of cement powder C-43 hydrated in water vapor on lowering RH (WR) (a), followed by wetting and drying (b) and grinding (c)

The quantity of P indicates the extent of hydration, whereas the quantity of A, V and calcite shows the sensitivity to carbonation. Main calcite peak coincides with that of alite, which is lowered on hydration, whereas calcite peak increases on carbonation and thus only the resultant may be analyzed (Figs 4d and 5d).

The SEM micrographs of hydrated cement powder are presented in Figs 6 and 7. The superficial hydration products as studied by TEM, are indicated in Figs 8–10.

*Comparison of hydration in water vapor and in liquid water as studied by static heating (Table 1)*

The content of all the hydration products, represented both by EV and non-EV, was higher in pastes hydrated in liquid water (by a factor of 1.5–1.7), than in powder hydrated in water vapor, due to the better availability of water molecules from liquid phase (Table 1a). Nevertheless the hydration process in the liquid water (pastes) was not terminated and proceeded in water vapor (both in WS and WR), increasing all the values of EV, of non-EV, of $\Delta M(220°C)$ and of $\Delta M(400°C)$. The increase in the last

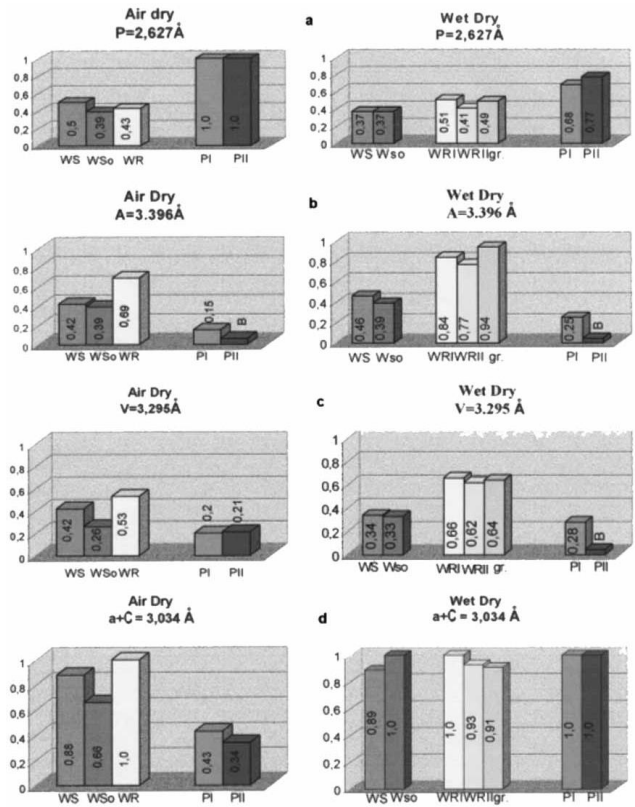

**Fig. 4** The influence of hydration in water vapor (WS, WSo and WR) and in liquid water (PI and PII) on the formation of (a) P – portlandite, (b) A – aragonite, (c) V – vaterite and (d) alite+calcite (a+C), in the air dry and wet dry state: histograms of the relative intensity (I) of the strongest XRD peaks of the products of cement hydration and carbonation, B – background

value (portlandite water content) was especially high and in series I it was in WS and WR double the value, measured in SH (carbonation ?, Table 1a, column g).

Powder samples did not show any rehydration effect after heating at 220°C and the content of hydration products was low. Rehydration in WS and WR was high in pastes, series I, amounting to about 10% as compared to $\Delta M$(220°C), Table 1a, compare column f and e. Also the mass after heating at 400°C increased as compared to that at 220°C. In the paste PI(WS) the rehydration caused the change from $\Delta M$(220°C)=15.6% to $\Delta M$(400°C)=16.4% and in paste PI(WR) from 14.4% to 18.0%, respectively, compare Table 1a, column e and g. The carbonation occurred thus at a temperature <400°C and also the sensitivity to carbonation was in this case high (Table 1a, column h). In cement powder (Table 1a) and in pastes, series II hydrated for 1 month (Table 1b) a systematic decrease in mass on heating was observed and rehydration was small, whereas in paste, series II hydrated for 6 months it was as high as in pastes, series I (1m), compare Table 1a and 1b, column f.

The content of P(H$_2$O), i.e. water chemically bound in portlandite, as estimated from $\Delta M(400°C)$ was increasing in the sequence:

powder<PI and PII, SH<series II, 1m, SH<series, II, 6m, SH

(4.5–4.8%)   (7.2–7.8%)        (8.9%)              (10.7%)

see Table 1, column g. Only in case of paste PI, the subsequent storage in water vapor caused its important increase, up to 16.4% (WS) and 18.0% (WR).

Surprisingly uniform values were obtained for $\Delta M(400°C)$ as measured in series II (6m) in all the three tests (SH, WS and WR), i.e. 10.5–10.8% (Table 1b, column g), though the difference in $\Delta M(600°C)$ was significant (0.8–6.0%, column h). This problem deserves farther study.

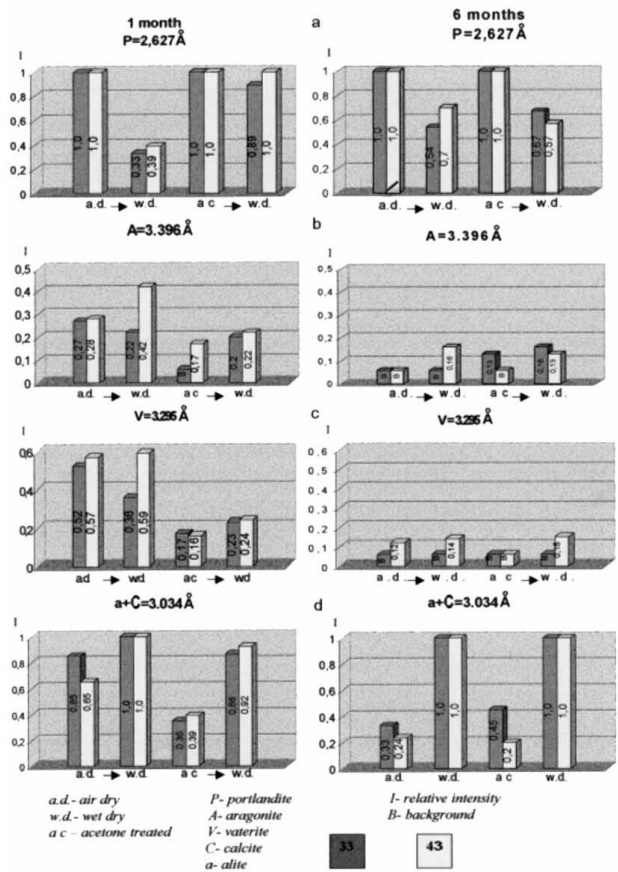

**Fig. 5** The influence of cement strength (C-33 and C-43 [31]) and hydration time (1 month and 6 months) on the formation of the products of hydration and carbonation in liquid phase: histograms of the relative intensity (I) of the strongest XRD peaks of (a) portlandite, P – 2.627 Å, (b) aragonite, A – 3.396 Å, (c) vaterite, V – 3.295 Å, (d) alite+calcite, a+C – 3.034 Å, B – background, a.d. – air dry, w.d. – wet dry, ac – acetone treated

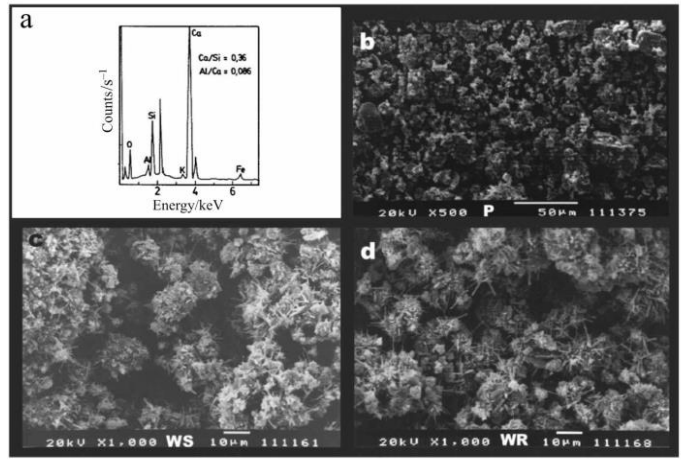

**Fig. 6** SEM micrographs of the cement powder (a) chemical composition as measured by EDX, (b) cement powder (375×500), (c) cement powder after hydration in water vapor at increasing RH, in WS (161×1000) and (d) at decreasing RH, in WR (168×1000)

The sensitivity to carbonation, characterized by the $\Delta M(600°C)$, was also lower in powder WS, than in WR and this was similar as in pastes PI and PII, SH (Table 1a, column h): 1.9% in WS, 3.0% in WR, and 2.7–3.4% in pastes of both series hydrated for 1 month. It was lowered to 0.8% after 6 months hydration.

Between the pastes series I (Table 1a) a series II (Table 1b) there are the following differences (Methods and footnote, Table 1): (*i*) age of cement powder differed by about 1 year, (*ii*) mixing and forming by hand of small cubes (series I) or standard procedure (series II), (*iii*) distilled water (series I) or tap water (series II). Thus the value of EV(SH) was lower in series II than in series I. Subsequent hydration in water vapor caused a slight increase in this value in WR, but a lowering in WS (Table 1b). The respective change in non-EV is rather small. Results of series II were discussed in a previous paper [31] and they were compared with the results obtained on a paste of a smaller strength.

*XRD study of portlandite and calcium carbonate*

XRD study was done about 1 year after hydration of both cement powder and pastes. The carbonates found by this method had supposedly formed during this time at room temperature on contact with air. Static heating was performed immediately after the termination of hydration process, thus it may be assumed that carbonate formation occurred only at the elevated temperature.

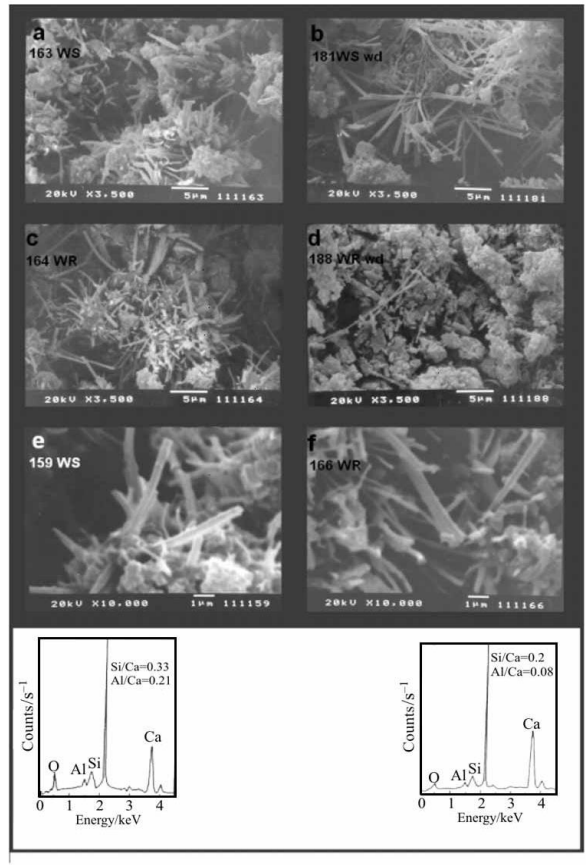

**Fig. 7** SEM micrographs of hydration products of cement powder (C-43): (a) after WS (163), (b) and wetting and drying (181), (c) after WR (164), (d) and wetting and drying (188) at ×3500; (e) after WS, Ca:Si=3.03, (159) (f) after WR, Ca:Si=5.0 (166), at ×10000

*Comparison of hydration in liquid distilled water and in water vapor (Fig. 4)*

The availability of water molecules was higher from liquid than from gaseous phase, thus the extent of hydration was also higher, compare the intensity of the main P peak at 2.63 Å (Fig. 4a) and the values of EV and non-EV in Table 1a and 1b.

The liquid phase and the dense microstructure, protected the paste samples from carbonation, the XRD peak intensity of calcite+alite, A and V is low (Fig. 4b–d); also the formation of big portlandite crystals in the paste may impede carbonation [31]. This reaction proceeded further on wetting and drying of the crushed samples, Figs 2b and 3b. The joint calcite+alite peak became the highest, both in the pastes and in cement powder, at the relative intensity, $I$=1.0. Also on heating of the samples of pastes PI and PII, and of cement powder, WR, a similar amount of carbonates was formed (Table 1a, column h), thus the sensitivity to carbonation was similar.

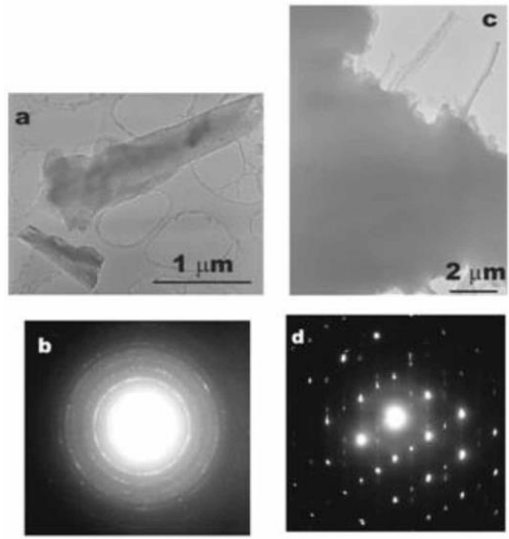

**Fig. 8** Comparison in TEM of various crystallinity in hydrated cement powder at a low magnification: (a) WR sample and (b) corresponding ED pattern, (c) WS sample and (d) corresponding ED pattern

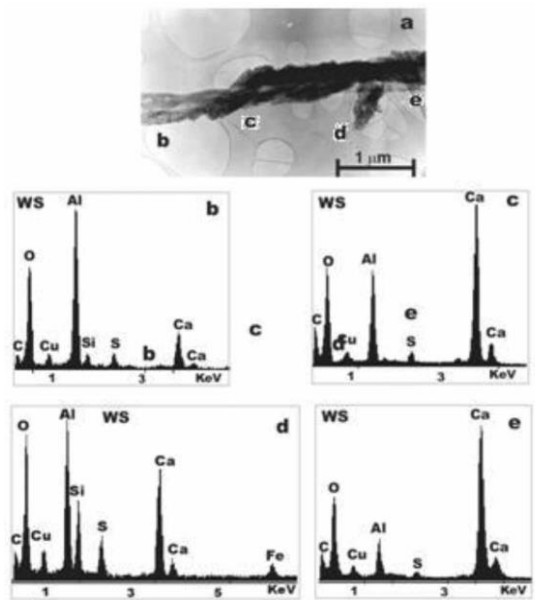

**Fig. 9** Migration of the ions along a needle: (a) TEM image and (b–e) EDX spectra corresponding to different parts of the needle. The small Al ions are concentrated at the left end (b) and in the small central needle (d). The big Ca ions are concentrated in the center (c) and in the right end of the needle (e)

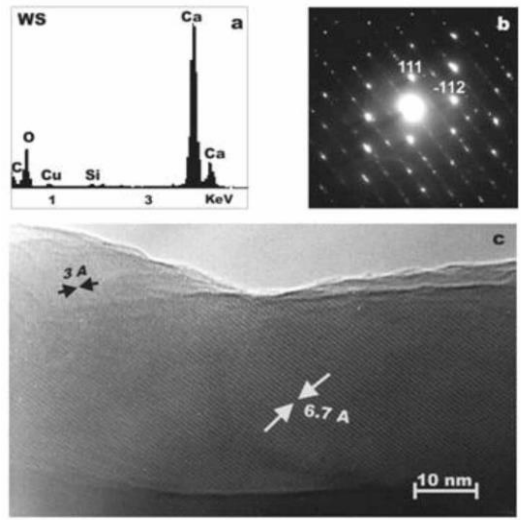

**Fig. 10a** Aragonite formation in the WS sample: (a) EDX spectrum, (b) ED pattern and (c) TEM micrograph at a high magnification, corresponding to an aragonite crystal (right) and a small area showing the possible vaterite formation (left)

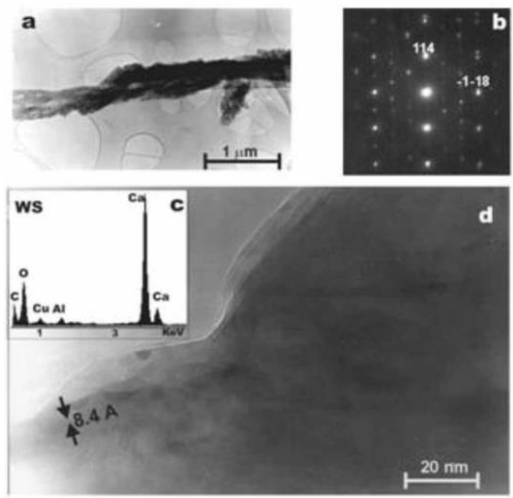

**Fig. 10b** Vaterite formation in the WS sample: (a) TEM image at a low magnification of a needle, Fig. 9a, (b) ED pattern corresponding to one end of the needle; the maxima can be indexed as vaterite, (c) EDX spectrum of the same area, (d) TEM image at a high magnification of vaterite

A small difference in the extent of hydration of cement powder was observed between WR and WS samples, where the P-peak was slightly higher, Fig. 4a (compare also $\Delta M$(400°C), Table 1a, column g), whereas in WR this compound transformed al-most completely into carbonates, Fig. 3a and the peak intensity was $I=0.69$, 0.53 and 1.0 for A, V and calcite+alite, respectively, Fig. 4b–d. In cement powder hydrated in WS the calcite+alite peak at 3.037 Å and $I=0.88$ was lower than in WR, whereas A and V indicated a similar peak intensity of $I=0.42$ (at 3.394 and 3.292 Å, respectively, Fig. 4) and the strongest was the joint alite+belite+vaterite peak at 2.75 Å and $I=1.0$.

Slight differences in basal spacing were observed, depending on sample preparation. The main (010) P-peak, nominally at 2.627 Å, was moved to 2.64 Å in cement powder and it was slightly higher in WS than in WR (compare the respective values of $\Delta M(400°C)$, Table 1a, column g). The (001) P-peak was in paste samples at the nominal spacing of 4.923 Å, whereas in powder samples it was displaced to 4.864 Å. Some coinciding component could be present, possibly analcime of the peak at 4.842 Å. In the WR sample this peak was found after wet-drying, Fig. 3b. Other peaks at a low intensity were also observed after this operation, and all these new peaks may correspond to portlandite and/or analcime. The shift of the P-peaks was not due to misalignment of the apparatus, as in all the samples the main calcite peak was found at the nominal spacing of 3.03–3.04 Å (Figs 2 and 3).

Little change in carbonation occurred in WS sample after its wetting and drying: no change in the calcite+alite peak intensity was observed, the A content increased only slightly, whereas the peaks of P and V were lowered, probably due to dissolution. In WR sample the carbonation proceeded after this operation and more A and V were found as their peak intensity increased, Fig. 4b and 4c (calcite+alite peak remained unchanged).

Very small amount of carbonates was found in paste samples PI and PII by XRD, whereas they formed at an elevated temperature and the sensitivity to carbon- ation was found similar in powder WR and paste PI and PII (Table 1a, column h).

This value was found the highest after WR test: the joint calcite+alite peak at 3.03 Å was the strongest ($I=1.0$, Fig. 4d) and also the A and V amount was appreciable ($I=0.69$ and 0.53, respectively, compare Table 1a, column h). In the WS sample the highest was the joint (alite+belite+vaterite) peak at 2.75 Å, the calcite+alite peak was also strong ($I=0.88$), whereas A and V indicated a similar peak intensity at $I=0.42$.

The grinding of the WR sample after its repeated wet-drying (WRI and WRII) caused a slight lowering of the calcite peak intensity and an increasing in that of the A-peak (Fig. 3c), (compare the influence of grinding of calcite [35]). The strongest P-peak at 2.64 Å remained unchanged but the peak at 4.85 Å (P and probably analcime) disappeared. The joint alite+belite+vaterite peak at 2.75 Å became the strongest. Thus grinding, applied for sample preparation may change the XRD trace.

Predrying of cement powder at 110°C (WSo) caused a lowering of its reactivity and the relative intensity of all the peaks was lowered, as compared to the WS sample (Fig. 4).

The start of precipitation of A from a supersaturated carbonate solution occurs at 40°C [18], though Bonatti et al. [36] report its crystallization from sea water at 1–5°C [23]. Thus it is a reaction dependent on temperature, on time and on solubility. High amount of A, which formed on WR at 30°C, indicates that calcium hydroxide, which formed on this hydration method is highly reactive, as it may be composed of small crystallites of a high specific surface (see SEM and TEM).

*Hydration and carbonation of the pastes in tap water (Fig. 5)*

The extent of hydration of the pastes series II was much higher after 6 months hydration than after 1 month: the alite peak intensity (joint with calcite) of the air dry samples was considerably lower in the first case (a+C, Fig. 5d, compare Table 1b, column g). The P-peak intensity was in both cases I=1.0. On the contrary, the extent of carbonation was much lower after 6 months hydration and the intensity of all the carbonate peaks was lower than that after 1 month hydration (Fig. 5b-d, i.e. A, V and C+a). Wet-drying induced carbonation in all the air dry samples. Acetone treatment prevented this reaction and the P-peak intensity was here I=1.0 also after wet-drying (Fig. 5.a). Carbonation occured though in acetone treated and wet-dried samples, both hydrated for 1 month and for 6 months (especially to calcite).

Some differences were observed between the pastes of different strength (C-33 and C-43), which was discussed in the previous paper [31].

*SEM and HRTEM study*

General remarks

The microstructure of the hydrated cement is being studied extensively by SEM (e.g. [5, 37]) and also by TEM, using sophisticated methods of sample preparation. Argon-ion milling and a liquid-nitrogen-cooled sample holder were applied to get a transparent specimen and to avoid radiation damage, which was observed on prolonged exposure, especially in sulfoaluminates [38]. Rayment and Majumdar [39] showed that the Ca count rates generally increased and Si count rates decreased with irradiation time: the Ca/Si ratio was 1.94 and 1.96 after 10 and 100 s measurement, respectively, increasing up to 20% after a longer irradiation time. According to Lachowski *et al.* [40] the apparent Ca/Si ratio was dependent on the beam voltage resulting in values 1.44 and 1.62 for 6 and 15 kV, respectively.

Very little radiation damage was observed during the present TEM study of the smallest particles protruding from grain surfaces of cement powder hydrated either in WS or in WR (SEM, Figs 6 and 7). These surface formations indicated two different shapes (HRTEM): either an elongated 'needle' or an irregular 'cluster'. They were studied at a lower magnification (ca. 10kx), at a high magnification (ca. 100kx), by electron diffraction (ED) and by energy dispersive X-ray analysis (EDX). The length of the needles was 2–14 μm and the width 0.2–0.6 μm; similar ones were observed in the pores of cement paste. The cluster size was about 0.5 μm and they contained a polycrystalline material. Well formed planes of a larger extension could only exceptionally be found. The distribution of chemical components was non-uniform.

More 'needles' were protruding from the surfaces of grains hydrated in WR than from WS samples [Fig. 6d (WR) as compared to Fig. 6c (WS) and Fig. 7c and f (WR), as compared to Fig. 7a, b and e]. This corresponds with the higher specific surface (proportional to EV, Table 1a, column c) and the higher reactivity shown by the net content of hydration products, i.e. non-EV, in the WR sample (Table 1a, column d). This higher reactivity caused also the almost complete transformation into carbonate after wetting and drying, seen as spongy white material in Fig. 7d. Also their chemical composition was different (Fig. 7, discussed below).

Polycrystallinity

Nanosized crystalline units with crystal planes of a varying spacing embedded in the amorphous matrix were found in the small particles (TEM, high magnification, see below).

The crystallinity was higher in WS samples than in the WR ones (Fig. 8), which may be due to a higher suction, exerted in the first case. Also a higher sensitivity to carbonation was observed in the second case (Table 1A, column h, Fig. 7d) but the XRD peaks were better formed after WS treatment (higher intensity, compare Figs 2 and 3).

Portlandite was observed in two forms: in WS sample a polycrystalline material, composed of nanosized crystals, was seen in the rim of a cluster with an amorphous center. In WR sample a whole cluster was composed entirely of nanosized portlandite crystals [31]. Popoola *et al*. [41] and Bortzmeyer *et al*. [42] observed similar nanosized hydration products in the polymer matrix of the macro-defect free cements and in composites of calcium aluminate cements with polyvinyl alcohol and of calcium silicate cements with polyacrylamide, Al being the main cross-linking ion in the polymer. Nanosized crystallites were supposed to play an important part in the mechanical properties of these cements.

Clusters were non-uniform, the cluster center being amorphous and containing mainly calcium and an increased amount of silica (CSH-gel), whereas the polycrystalline material was formed on the cluster edges in small units (about 10 nm) and of varying chemical composition. Microcrystals of calcite were observed within the outer product only, accompanied by $Ca(OH)_2$ as much larger imperfect crystals [31]. Groves *et al*. [43] found that the carbonated paste (1 year) consisted of small residual particles embedded in an extremely homogeneous gel as inner product.

Variation in chemical composition

The chemical composition as measured in hydrated cement by EDX in SEM was consistent with that measured in cement powder by standard methods (compare Materials). The differences in content of CaO and $SiO_2$ are due to the selection of a P-free surface, as studied by SEM.

The chemical composition of the surface formations of the hydrated cement was dependent on hydration conditions (Fig. 7). The Ca content measured by EDX (both in SEM and TEM) was higher in WR sample. This is analogous to the influence of irradiation, see above [39]. Al accumulated on the surfaces of WS samples, usually accompanied by a small amount of S (ettringite), Figs 7e and f.

In the central needle (WR), the ratio of Ca:Si was equal to 5 (point acquisition, Fig. 7b), whereas in WS it was 3.03 (Fig. 7). Increased content of Ca was generally measured also on other surfaces of WR samples, both by SEM and TEM. This caused a higher sensitivity to carbonation; a higher amount of carbonates formed in this sample and was identified by XRD (Figs 3 and 4, compare [44, 45]).

An uneven distribution of some ions indicates the possibility of dissolution and diffusion along the needle: the smaller Al ions ($r$=0.039 nm) were accumulated on the left hand side, Fig. 9a, b and d, diffusing more quickly than the bigger Ca-ions ($r$=0.10 nm), which were concentrated at the right hand side of this needle (Fig. 9c and e, WS sample). The possibility of the dissolution-diffusion-crystallization pro- cess was suggested by Yariv and Cross [46]. Dissolution could have also contributed to the increased sensitivity to carbonation, observed in paste samples (6m) after sub- sequent sorption of water vapor in WS (Table 1b, column h).

Along another needle a decrease in Ca content and increase in Al and Si content was observed, their ratio of Ca/(Al+Si) being 3/1 at one end, 2/1 at the center and 1/3 at the other end. The average is in agreement with the value indicated by Richardson and Groves [44], i.e. Ca/Si=1.8±0.4. Some small sulfur content showed the presence of ettringite at the one end of the needle (compare Fig. 9).

Richardson and Groves [44] found that the Ca/Si ratio in the CSH-gel ($nCaO \cdot mSiO_2 \cdot pH_2O$) reveals a significant variation from one region to another and it varies also significantly with $w/c$ ratio. Some Ca may be replaced by Mg whereas Al, S and Fe may replace Si [39]. Less CH, $[Ca(OH)_2]$, was found in CSH-gel at a lower $w/c$ ratio, but more high Ca determinations were obtained in this case, either because of unhydrated $C_3S$ ($3CaO \cdot SiO_2$, alite) or because of the presence of microcrystalline CH [47].

In the superficial formation studied here, the Si content was generally much smaller than Ca content. Exceptionally in one cluster equal amounts of Ca, Si and O were de- tected and also some Al, K, Fe, Na and Cl were identified in this place, which looks like a 'waste repository of surplus ions'. This suggests that the hydration process includes the dissolution stage and crystallization of a specific compound (compare [46]).

EDX spectra at a point acquisition indicate an uneven distribution of particular com- pounds in cement powder itself, which could contribute to the formation of aragon- ite [26], enhanced also by the temperature of drying after wetting (w.d., 40°C, [18]). The ratio of Ca/Si in unhydrated cement powder (500×), was 3.2 to 2.7 and it was 3.2 to 3.1 in hydrated paste. On point acquisition and at a high magnification it varied between 5.0 and 1.8 in hydrated powder and 10 to 2.1 in hydrated paste. Lachowski and Diamonds [45] observed also variations within micrometer sized fragments, Ca/Si changing between 0.89 to over 3, mostly 1.3 to 1.8. This is attributed to the difficulty in ionic transport within the paste, thus the metastable equilibrium may be attained only locally on mi- crometer or even smaller scale [38]: important compositional variations were found be- tween the particles of CSH-gel (calcium silicate hydrate) and sulphoaluminate phases.

Vaterite and aragonite in WS and WR samples

ED in HRTEM study of the WR powder samples indicated most frequently portlandite, containing some Si and Al. It was less abundant in WS samples, probably hidden in the sample interior.

In WS sample several zones of aragonite and vaterite structure were identified, showing extensive crystal planes (Figs 10a and 10b). They were free of Si and Al and their crystal planes were well formed. Apparently storing for a prolonged time at a low and constant RH produced some ordering of microstructure which induced an easier crystallization.

Only one aragonite zone was found in the WR sample and here also some Si and Al were present.

## Conclusions

Cement hydration products were studied as influenced by the hydration conditions: (*i*) hydration time in liquid phase, either for 1 month (1m) or for 6 months (6m); (*ii*) cement 'age': pastes were prepared in series I and about 1 year later in series II, (*iii*) relative humidity, RH, in gaseous phase, either increasing in water sorption WS (RH=$0.5 \rightarrow 0.95 \rightarrow 1.0$), or being lowered in water retention, WR (RH=$1.0 \rightarrow 0.95 \rightarrow 0.5$). Either distilled or tap water was used and the test cubes were formed either by hand or by standard method. The formation of calcium hydroxide (portlandite, P) and its transformation to calcium carbonates is mainly discussed here.

*1)* More hydration products, including P, were formed in liquid phase (paste) than in water vapor (powder), due to the higher availability of water molecules.

*2)* Full hydration was observed only in the paste (6m), otherwise the P content, as estimated from its water escape, $\Delta M(400–800°C)$, increased after storage of the prehydrated paste in water vapor. In the paste (6m) this value did not change after WS or WR and in all cases it was amounting to 10.5–10.8%.

*3)* The extent of carbonation, proceeding in time at room temperature in air (XRD), was higher in cement hydrated from gaseous phase (than from liquid phase), as its microstructure is less compact, thus the $CO_2$ from air is easier accessible. Also the portlandite crystals may be smaller. All the three $CaCO_3$ polymorphs were forming in hydrated powder, i.e. calcite, aragonite and vaterite. Their content was higher on hydration in WR than in WS. Wetting and drying caused an increase in the content of all the three polymorphs in WR sample, whereas in WS sample only a slight change was observed (slight decrease in V content).

*4)* Accumulation of Ca on the surfaces of cement grains hydrated in WR was observed in SEM and TEM and this provoked the carbonation. On the microsurfaces of WS samples the Al content was increased.

*5)* The content of carbonate polymorphs decreased in the order: calcite>V>A. After wetting and drying only the content of calcite was higher (V content was somewhat lowered). Acetone treatment prevented the formation of carbonates.

*6)* The $\Delta M(600–800°C)$ was assumed to represent the sensitivity to carbonation, i.e. the decomposition of $CaCO_3$, which formed from portlandite on heating between 400 and 600°C and decomposed at 600–800°C. It was similar in pastes and in powder hydrated in WR. It was lower in WS and much lower in paste hydrated for 6 months. It increased pronouncedly after the subsequent hydration in water vapor in WS of the prehydrated pastes; no change was found in WR in the 'younger' paste (series I, 1m). In the 'older' paste (series II, 1m) this value was close to zero and in series II (6m) it increased pronouncedly similarly as in series I.

*7)* Migration of ions was observed along the elongated 'needles' on the surfaces of the hydrated powder grains (TEM). It was dependent on the ion size (small Al-ions were concentrated at one end and bigger Ca-ions – at the other end of some needles), indicating dissolution-diffusion-recrystallization process. The inner product in clusters was composed of amorphous CSH-gel; whereas on the rims the Ca ions were accumulated, forming a polycrystalline material (TEM).

*8)* Hydration conditions influence the reactivity of the produced calcium hydroxide, which reactivity is dependent on the size of its crystallites and/or on their crystallinity. The crystallinity of the hydration products was higher after WS than after WR (XRD, ED). In WS some stress seems to be exerted and fixation of Ca ions occurs. In both cases nanosized crystallites were found in the amorphous matrix, which could contribute to the strength of the hydrated cement.

* * *

This study was partly done within an exchange visitor program between the Polish Academy of Sciences, the 'Consejo Superior de Investigaciones Cientificas' and the Indian National Science Academy. The cement powder and paste samples, series II, were supplied by Dr A. K. Mullick and Dr S. K. Handoo from the National Council for Cement and Building Materials, New Delhi, India. Sorption tests and heating tests were done by Ms Krystyna Wiellowicz at IBW PAN, Gdańsk. XRD study on carefully crushed samples was performed by Dr M.A. Aviles and SEM study was done by Ms M. C. Jiménes de Haro, both at ICMS CSIC, Sevilla, Spain. Partial financial support was obtained from the project MAT 1999-0995.

## Appendix

*Definitions of the symbols:*
*Compounds:*

$C_3S=3CaO \cdot SiO_2=Ca_3SiO_5$, alite
$C_2S=2CaO \cdot SiO_2=Ca_2SiO_4$, belite (larnite)
$CH=Ca(OH)_2$
$P=Ca(OH)_2$, portlandite
$P(H_2O)$=content of water, bound chemically in portlandite
$A=CaCO_3$, aragonite
$V=CaCO_3$, vaterite
CSH-gel=$[nCaO \cdot mSiO_2 \cdot pH_2O]$

*Methods:*

WS – water sorption at the relative intensity increasing successively: RH=0.5$\rightarrow$0.95$\rightarrow$1.0
WSo – water sorption of cement powder predried at 110°C
WR – water retention at decreasing successively RH=1.0$\rightarrow$0.95$\rightarrow$0.5
PI and PII – paste samples series I, mixed by hand at distilled water/cement ratio $w/c$=0.4; cubes 3×3×3 cm, hydrated for 28 days
paste C-43 - series II, prepared at tap water/cement ratio $w/c$=0.4; standard cubes hydrated 1m – 1 month, 6m – 6 months, a.d. – air dried, ac – acetone treated and air dried

SH – static heating of hydrated cement paste

*I* – relative intensity of the XRD peaks, indicated by the computer, related to the highest peak intensity of *I*=1.0

*Hydration products:*

EV=water content (W) escaping at 110°C

$\Delta M(T)$ – mass change on heating at the given temperature T°C, related to the mass at 800°C, i.e. $\Delta M(800°C)=0$ and non-EV=$\Delta M(110°C)$

$\Delta M(110–220°C)$ – content of water, bound with low energy

$\Delta M(220–400°C)$ – content of water, bound in the poorly crystalline CSH-gel

$\Delta M(400–800°C)=P(H_2O)$ – content of water, released from portlandite on its dehydroxylation

$\Delta M(600–800°C)$ – sensitivity to carbonation, i.e. $CO_2$ escaping within this temperature range, bound previously at a lower temperature.